        \documentclass[useAMS]{mn2e}
        \usepackage{graphicx}

\def\oversim#1#2{\lower0.5pt\vbox{\baselineskip0pt \lineskip-0.5pt
     \ialign{$\mathsurround0pt #1\hfil##\hfil$\crcr#2\crcr\sim\crcr}}}
\title[The 2003 Shell Event in $\eta$ Carinae]
{The 2003 Shell Event in $\eta$ Carinae\thanks{This
paper is based on observations made at the South African
Astronomical Observatory.}}

\author[P.A. Whitelock et al.]
{Patricia A. Whitelock$^1$\thanks{e-mail: \tt paw@saao.ac.za}, 
    Michael W. Feast$^2$, Freddy Marang$^1$, Elm\'e Breedt$^3$\\
      $^1$ South African Astronomical Observatory, P.O.Box 9, 7935
           Observatory, South Africa.\\
       $^2$ Astronomy Department, University of Cape Town, 7701 Rondebosch,
           South Africa.\\
       $^3$ National Astrophysics and Space Science Programme,
            University of Cape Town, 7701 Rondebosch, South Africa.\\
}
\date{Received date; accepted date}

\pubyear{2004}
\begin{document}
\maketitle

\begin{abstract} Near-infrared, $JHKL$, photometry of $\eta$~Car is reported
covering the period 2000 to 2004. This includes the 2003 shell event which
was the subject of an international multi-wavelength campaign. The fading
that accompanied this event was similar to, although slightly deeper than,
that which accompanied the previous one. The period between these events is
$2023\pm3$ days and they are strictly periodic. Their cause, as well as that
of the quasi-periodic variations and secular brightening are discussed. It
seems possible that all three types of variability are consequences of the
binary nature of the star.
\end{abstract}


\section{Introduction}
 $\eta$ Carinae is one of the most luminous stars in the Galaxy, and despite
being one of the best studied objects in the sky\footnote{As of February
2004 NASA's ADS lists 844 publications mentioning $\eta$ Car in the
abstract, 250 of them published since 2000.0.}, it remains poorly
understood. While most of its luminosity is emitted in the mid-infrared,
there is flux from and variability at all wavelengths from hard X-rays to
radio. $\eta$ Car is in some senses the archetypal astronomical object -
which can only be fully characterized through observations across the entire
electromagnetic spectrum. Davidson \& Humphreys (1997) provide an extensive
review of $\eta$ Car, though there have been many developments since its
completion\footnote{see, e.g., http://etacar.umn.edu.}.

It seems likely that we will soon be observing the details of similar 
objects in distant galaxies, so an understanding of this object
potentially bears on many other topics. 

Damineli (1996) discovered periodic variations in the equivalent width of
HeI 1.083\,$\mu$m, and in the visibility of high excitation lines, which
correlated with $JHK$ flux variations. He consequently suggested that $\eta$
Car was in fact a binary system with an orbital period of about 5.5 years. 
Damineli predicted the next ``shell phase" for 1997/98 and $\eta$ Car was
subjected to a period of intensified observations at all wavelengths (Morse
et al. 1999). Not only did the ``shell phase" occur on schedule, but it was
accompanied by a number of unexpected phenomena, the most dramatic of which
was a very sharply defined X-ray minimum (Corcoran et al. 2001). This was
associated with, among other things, a very clear dip in the $JHKL$ light
curves (Whitelock \& Laney 1999), which we refer to as the ``eclipse-like
event'' or simply the ``event'' in the following. Subsequent detailed
examination of historical spectra (Feast et al. 2001) and monitoring at
various wavelengths (e.g. 3\,cm flux - Duncan \& White 2003) established
that a number of phenomena varied on this same timescale of 5.5 years.

The 5.5-year periodicity is now well established and most observers are
convinced that $\eta$ Car is a binary system. Some doubts do remain on the
binary issue because, to date, none of the orbital parameter determinations
have been convincing and there seem to be more questions about the
underlying source(s) than ever before. The most recent ``shell phase"
occurred during 2003 and was observed even more intensively than the
previous one. This paper is our contribution to the understanding of that
event and we report broad-band near-infrared $JHKL$ photometry obtained over
the last 5 years. It continues the series of papers which collectively cover
32 years of photometry of $\eta$ Car (Whitelock et al. 1983 \& 1994; Feast
et al. 2001 - henceforth Papers I, II and III). We start with a detailed
discussion in section 3 of the source of the near-infrared flux as it is
somewhat different from that of the visible, mid-infrared or far-infrared
radiation. It is crucial that we have some clarity on this before we go on
to consider the variations in the $JHKL$ light curves and their various time
scales in section 4.

In discussing the 5.5-year periodic phenomena, of which the 2003 shell event
is the latest example, we refer our timings to the X-ray minimum observed by
Corcoran et al. (2001) and Corcoran (2004). In particular we use Julian Day
2452819.7 (2003.49) as phase zero; this coincides with the start of the
minimum (the faintest point) of the X-ray ``eclipse''. In doing this we make
no particular comment on the correctness or otherwise of the Corcoran et al.
model which has periastron at phase zero. The 2022 day period derived by
Corcoran et al. is essentially identical to the 2023 days we derive from the
$JHKL$ light curves (section 4.2).

In Paper III we interpreted the dip in the infrared light-curve, which
repeats every 5.5 yrs, as an eclipse of a hot spot. We now believe that the
balance of evidence does not favour what would normally be called an
eclipse, as is discussed below in detail. Nevertheless, we continue to use
terminology that is usually associated with eclipses (e.g. ``fourth
contact'' in the caption to Fig.~5) simply because it provides a succinct
description of what we observe. It is important that the reader is clear
that the use of this terminology does not imply the existence of eclipses.

This may well be the last paper in this series as we anticipate the 
closure of the infrared photometer on the 0.75m telescope at Sutherland 
within the next year.

%
% TABLE 1
%
\begin{table}
\begin{center}
\caption{Near-Infrared Photometry}
\begin{tabular}{rrrrr}
\hline
\multicolumn{1}{c}{JD}&\multicolumn{1}{c}{$J$}&
\multicolumn{1}{c}{$H$}&\multicolumn{1}{c}{$K$}&\multicolumn{1}{c}{$L$}\\
\multicolumn{1}{c}{--2450000}&\multicolumn{4}{c}{(mag)}\\
\hline
1709.00 &  2.490 &  1.599 &  0.454 & --1.708\\
1712.22 &  2.487 &  1.597 &  0.450 & --1.696\\
1714.25 &  2.474 &  1.598 &  0.449 & --1.702\\
1737.23 &  2.499 &  1.615 &  0.466 & --1.685\\
1738.21 &  2.507 &  1.619 &  0.463 & --1.694\\
1743.21 &  2.514 &  1.607 &  0.458 & --1.681\\
1808.66 &  2.543 &  1.643 &  0.479 &  \\
1834.63 &  2.551 &  1.627 &  0.473 & --1.668\\
1858.61 &  2.543 &  1.651 &  0.480 & --1.725\\
\vspace*{2mm}
1862.61 &  2.547 &  1.643 &  0.483 & --1.692\\
1869.62 &  2.508 &  1.619 &  0.470 & --1.723\\
1881.61 &  2.533 &  1.620 &  0.453 & --1.705\\
1890.62 &  2.518 &  1.614 &  0.448 & --1.730\\
1926.56 &  2.572 &  1.672 &  0.498 & --1.740\\
1929.54 &  2.574 &  1.683 &  0.504 & --1.717\\
1962.47 &  2.571 &  1.664 &  0.495 & --1.687\\
1965.45 &  2.559 &  1.670 &  0.489 & --1.734\\
1976.45 &  2.580 &  1.666 &  0.481 & --1.710\\
2032.34 &  2.487 &  1.586 &  0.448 & --1.758\\
\vspace*{2mm}
2037.29 &  2.479 &  1.591 &  0.438 & --1.773\\
2068.24 &  2.462 &  1.560 &  0.418 & --1.717\\
2087.24 &  2.476 &  1.571 &  0.422 & --1.732\\
2093.25 &  2.492 &  1.577 &  0.428 & --1.715\\
2116.19 &  2.440 &  1.531 &  0.384 & --1.759\\
2211.62 &  2.454 &  1.538 &  0.393 & --1.729\\
2239.61 &  2.422 &  1.525 &  0.409 & --1.733\\
2257.60 &  2.407 &  1.518 &  0.384 & --1.767\\
2261.62 &  2.454 &  1.534 &  0.390 & --1.716\\
2267.62 &  2.394 &  1.529 &  0.391 & --1.783\\
\vspace*{2mm}
2283.58 &  2.426 &  1.512 &  0.371 & --1.750\\
2284.61 &  2.464 &  1.551 &  0.405 & --1.710\\
2285.58 &  2.456 &  1.551 &  0.410 & --1.727\\
2287.57 &  2.456 &  1.547 &  0.411 & --1.751\\
2288.59 &  2.489 &  1.586 &  0.466 & --1.693\\
2289.64 &  2.466 &  1.562 &  0.426 & --1.725\\
2318.52 &  2.468 &  1.568 &  0.419 & --1.738\\
2320.49 &  2.430 &  1.538 &  0.399 & --1.752\\
2324.49 &  2.462 &  1.575 &  0.427 & --1.760\\
2346.43 &  2.474 &  1.555 &  0.415 & --1.747\\
\vspace*{2mm}
2350.43 &  2.464 &  1.558 &  0.411 & --1.759\\
2352.40 &  2.452 &  1.551 &  0.402 & --1.776\\
2381.37 &  2.451 &  1.549 &  0.409 & --1.737\\
2386.33 &  2.461 &  1.553 &  0.411 & --1.744\\
2421.25 &  2.469 &  1.544 &  0.406 & --1.705\\
2426.23 &  2.463 &  1.553 &  0.408 & --1.731\\
2428.27 &  2.472 &  1.558 &  0.412 & --1.715\\
2451.26 &  2.460 &  1.539 &  0.406 & --1.702\\
2455.24 &  2.467 &  1.550 &  0.406 & --1.726\\
2486.23 &  2.514 &  1.585 &  0.427 & --1.691\\
\vspace*{2mm}
2490.19 &  2.512 &  1.582 &  0.420 & --1.695\\
2492.20 &  2.513 &  1.587 &  0.421 & --1.692\\
2525.68 &  2.487 &  1.561 &  0.397 & --1.711\\
2529.67 &  2.512 &  1.580 &  0.413 &  \\
2534.65 &  2.495 &  1.591 &  0.431 & --1.717\\
2563.64 &  2.533 &  1.577 &  0.420 &  \\
2564.63 &  2.488 &  1.585 &  0.419 & --1.725\\
2568.64 &  2.478 &  1.557 &  0.399 & --1.747\\
2573.63 &  2.473 &  1.566 &  0.412 & --1.744\\
2593.62 &  2.490 &  1.563 &  0.413 & --1.752\\
2603.60 &  2.479 &  1.562 &  0.411 & --1.737\\
\hline
\end{tabular}
\end{center}
\end{table}
\setcounter{table}{0}
\begin{table}
\begin{center}
%\caption[]{continued}
\begin{tabular}{rrrrr}
\hline
\multicolumn{1}{c}{JD}&\multicolumn{1}{c}{$J$}&
\multicolumn{1}{c}{$H$}&\multicolumn{1}{c}{$K$}&\multicolumn{1}{c}{$L$}\\
\multicolumn{1}{c}{--2450000}&\multicolumn{4}{c}{(mag)}\\
\hline
2661.57 &  2.490 &  1.554 &  0.382 & --1.761\\
2664.57 &  2.485 &  1.560 &  0.387 & --1.783\\
2665.59 &  2.481 &  1.550 &  0.394 & --1.770\\
2667.56 &  2.480 &  1.554 &  0.390 & --1.781\\
2675.56 &  2.483 &  1.557 &  0.393 & --1.761\\
2676.51 &  2.483 &  1.562 &  0.386 & --1.783\\
2680.52 &  2.512 &  1.560 &  0.377 & --1.747\\
2689.49 &  2.481 &  1.561 &  0.404 & --1.760\\
2690.44 &  2.497 &  1.548 &  0.399 & --1.749\\
\vspace*{2mm}
2691.49 &  2.499 &  1.563 &  0.399 & --1.755\\
2692.48 &  2.483 &  1.552 &  0.390 & --1.745\\
2693.52 &  2.482 &  1.553 &  0.389 & --1.759\\
2694.50 &  2.487 &  1.551 &  0.385 & --1.746\\
2695.48 &  2.484 &  1.553 &  0.396 & --1.757\\
2696.55 &  2.477 &  1.554 &  0.396 & --1.749\\
2697.54 &  2.474 &  1.555 &  0.391 & --1.752\\
2698.50 &  2.503 &  1.548 &  0.384 & --1.723\\
2699.47 &  2.478 &  1.547 &  0.388 & --1.757\\
2700.45 &  2.504 &  1.546 &  0.381 & --1.719\\
\vspace*{2mm}
2701.52 &  2.481 &  1.539 &  0.389 & --1.751\\
2702.46 &  2.489 &  1.542 &  0.384 & --1.727\\
2710.44 &  2.476 &  1.556 &  0.383 & --1.747\\
2711.49 &  2.477 &  1.540 &  0.378 & --1.756\\
2712.54 &  2.462 &  1.529 &  0.368 & --1.757\\
2715.53 &  2.488 &  1.529 &  0.327 & --1.784\\
2716.43 &  2.455 &  1.511 &  0.359 & --1.785\\
2752.34 &  2.426 &  1.454 &  0.298 & --1.806\\
2753.33 &  2.428 &  1.458 &  0.300 & --1.793\\
2754.31 &  2.402 &  1.452 &  0.294 & --1.806\\
\vspace*{2mm}
2755.32 &  2.406 &  1.448 &  0.298 & --1.807\\
2756.30 &  2.406 &  1.434 &  0.290 & --1.806\\
2758.31 &  2.376 &  1.429 &  0.278 & --1.830\\
2759.36 &  2.399 &  1.423 &  0.282 & --1.816\\
2760.29 &  2.397 &  1.436 &  0.280 &  \\
2761.29 &  2.384 &  1.424 &  0.282 & --1.806\\
2762.29 &  2.396 &  1.428 &  0.283 & --1.814\\
2764.42 &  2.382 &  1.430 &  0.273 & --1.809\\
2773.39 &  2.355 &  1.400 &  0.270 & --1.825\\
2774.26 &  2.324 &  1.392 &  0.256 & --1.842\\
\vspace*{2mm}
2775.28 &  2.344 &  1.392 &  0.257 & --1.850\\
2777.23 &  2.337 &  1.386 &  0.256 & --1.840\\
2778.27 &  2.339 &  1.389 &  0.255 & --1.837\\
2779.25 &  2.335 &  1.388 &  0.254 & --1.856\\
2794.24 &  2.339 &  1.360 &  0.240 & --1.812\\
2795.21 &  2.339 &  1.367 &  0.239 & --1.826\\
2797.22 &  2.327 &  1.351 &  0.237 & --1.820\\
2798.22 &  2.326 &  1.346 &  0.225 & --1.807\\
2800.22 &  2.324 &  1.349 &  0.227 & --1.824\\
2801.20 &  2.321 &  1.332 &  0.214 & --1.810\\
\vspace*{2mm}
2802.23 &  2.305 &  1.317 &  0.209 & --1.817\\
2803.23 &  2.311 &  1.338 &  0.222 & --1.827\\
2804.21 &  2.299 &  1.315 &  0.217 & --1.833\\
2805.21 &  2.293 &  1.319 &  0.211 & --1.831\\
2806.20 &  2.305 &  1.318 &  0.211 & --1.821\\
2807.20 &  2.292 &  1.307 &  0.208 & --1.829\\
2809.28 &  2.291 &  1.310 &  0.207 & --1.843\\
2810.27 &  2.291 &  1.301 &  0.200 & --1.815\\
2811.26 &  2.282 &  1.301 &  0.203 & --1.836\\
2813.30 &  2.283 &  1.308 &  0.221 & --1.800\\
2814.29 &  2.281 &  1.307 &  0.215 & --1.813\\
\hline
\end{tabular}
\end{center}
\end{table}
\setcounter{table}{0}
\begin{table}
\begin{center}
\caption[]{continued}
\begin{tabular}{rrrrr}
\hline
\multicolumn{1}{c}{JD}&\multicolumn{1}{c}{$J$}&
\multicolumn{1}{c}{$H$}&\multicolumn{1}{c}{$K$}&\multicolumn{1}{c}{$L$}\\
\multicolumn{1}{c}{--2450000}&\multicolumn{4}{c}{(mag)}\\
\hline
2815.22 &  2.287 &  1.312 &  0.217 & --1.785\\
2816.33 &  2.267 &  1.318 &  0.238 & --1.785\\
2817.25 &  2.289 &  1.318 &  0.239 & --1.773\\
2818.19 &  2.301 &  1.324 &  0.244 & --1.756\\
2819.27 &  2.300 &  1.328 &  0.249 & --1.748\\
2820.21 &  2.288 &  1.325 &  0.261 & --1.728\\
2821.26 &  2.307 &  1.340 &  0.274 & --1.726\\
2823.23 &  2.303 &  1.352 &  0.297 & --1.713\\
2825.19 &  2.332 &  1.384 &  0.324 & --1.695\\
\vspace*{2mm}
2826.16 &  2.349 &  1.400 &  0.333 & --1.669\\
2828.25 &  2.366 &  1.424 &  0.370 & --1.663\\
2831.21 &  2.403 &  1.450 &  0.400 & --1.581\\
2832.20 &  2.407 &  1.459 &  0.406 & --1.597\\
2833.21 &  2.414 &  1.472 &  0.413 & --1.558\\
2834.21 &  2.413 &  1.484 &  0.425 & --1.572\\
2835.19 &  2.420 &  1.485 &  0.430 & --1.585\\
2840.25 &  2.436 &  1.487 &  0.449 & --1.582\\
2841.25 &  2.440 &  1.484 &  0.441 & --1.560\\
2842.21 &  2.421 &  1.474 &  0.437 & --1.569\\
\vspace*{2mm}
2843.20 &  2.429 &  1.473 &  0.436 & --1.586\\
2844.20 &  2.423 &  1.469 &  0.437 & --1.560\\
2845.20 &  2.408 &  1.451 &  0.418 & --1.562\\
2846.19 &  2.406 &  1.459 &  0.433 & --1.561\\
2847.19 &  2.412 &  1.449 &  0.416 & --1.561\\
2848.19 &  2.399 &  1.438 &  0.407 & --1.570\\
2849.19 &  2.394 &  1.436 &  0.395 & --1.559\\
2850.19 &  2.363 &  1.431 &  0.398 & --1.585\\
2852.19 &  2.346 &  1.413 &  0.379 & --1.625\\
2854.19 &  2.354 &  1.411 &  0.384 & --1.595\\
\vspace*{2mm}
2855.19 &  2.341 &  1.402 &  0.367 & --1.617\\
2856.19 &  2.311 &  1.391 &  0.358 & --1.645\\
2857.19 &  2.353 &  1.394 &  0.363 & --1.610\\
2858.22 &  2.329 &  1.385 &  0.356 & --1.626\\
2865.19 &  2.300 &  1.369 &  0.342 & --1.653\\
2868.19 &  2.303 &  1.378 &  0.335 & --1.652\\
2869.19 &  2.302 &  1.375 &  0.330 & --1.671\\
2886.23 &  2.268 &  1.322 &  0.283 & --1.697\\
2887.24 &  2.261 &  1.331 &  0.286 & --1.703\\
2891.23 &  2.246 &  1.320 &  0.278 & --1.720\\
\vspace*{2mm}
2893.23 &  2.256 &  1.300 &  0.270 & --1.729\\
2894.22 &  2.261 &  1.326 &  0.269 & --1.704\\
2896.23 &  2.232 &  1.333 &  0.274 & --1.767\\
2918.64 &  2.240 &  1.319 &  0.272 & --1.721\\
2931.64 &  2.238 &  1.322 &  0.260 & --1.696\\
2932.63 &  2.252 &  1.329 &  0.265 & --1.691\\
2933.64 &  2.244 &  1.336 &  0.274 & --1.731\\
2955.62 &  2.265 &  1.351 &  0.298 & --1.727\\
2959.62 &  2.255 &  1.358 &  0.301 & --1.744\\
2961.61 &  2.271 &  1.362 &  0.294 & --1.718\\
\vspace*{2mm}
2969.61 &  2.264 &  1.359 &  0.290 & --1.723\\
2971.60 &  2.247 &  1.370 &  0.292 & --1.728\\
2972.61 &  2.269 &  1.365 &  0.305 & --1.705\\
2973.61 &  2.261 &  1.371 &  0.285 & --1.714\\
2974.61 &  2.266 &  1.386 &  0.290 & --1.716\\
2975.61 &  2.250 &  1.351 &  0.286 & --1.726\\
2978.62 &  2.269 &  1.367 &  0.312 & --1.756\\
3026.50 &  2.287 &  1.434 &  0.352 & --1.738\\
3029.60 &  2.300 &  1.435 &  0.357 & --1.731\\
3030.50 &  2.311 &  1.435 &  0.361 & --1.680\\
\hline
\end{tabular}
\end{center}
\end{table}

\section{Infrared Photometry}
 Monitoring of $\eta$ Car in $JHKL$ at SAAO in the period 1972-2000
established a quasi-cyclical variation with a period of the order of 5 years
superposed on a secular increase in brightness (see Papers~II and III).
Table 1 contains the results of a continuation of this series of
observations for the period 2000-2004. As in all recent work in this series,
the observations were made with the SAAO MkII photometer on the 0.75m
reflector at SAAO, Sutherland, using a 36\,arcsec diaphragm encompassing the
entire bipolar nebula (the Homunculus nebula). The individual results are
accurate to better than $\pm$0.03 mag in $JHK$ and $\pm$0.05 mag in $L$. The
magnitudes are in the system defined by the standard stars of Carter (1990).

 The magnitudes reported here, together with those reported in Papers~I to
III, are shown in Fig.~1 as a function of time.  The last two 
cycles are shown on an expanded time scale in Fig.~2. Fig.~\ref{jl}
illustrates how the $J-L$ colour of $\eta$~Car changes as a function of
time, while Fig.~\ref{jkkl} shows a two-colour diagram with the various
cycles identified by different symbols. There is a progression with
time from the top right to the bottom left as the star gets bluer in all
colours. The very first cycle has a lower $J-K$ than might have been
predicted from its $K-L$ and the later colours. A brightening or `flare' at
$J$ is seen during the second cycle, more specifically during the first few
months of 1977. This could have been caused by a strengthening of the
emission lines and the most likely culprit would be HeI 1.083\,$\mu$m (see
Whitelock (1985) for an indication of how this line affects the SAAO $J$
magnitudes).

The near-infrared spectrum has been discussed by various authors (Paper~I;
Allen, Jones \& Hyland 1985; McGregor, Hyland \& Hillier 1988; Hamann et al.
1994; Smith 2002). The nature of the near-infrared emission is examined in
detail below.

\begin{figure}
\includegraphics[width=8.5cm]{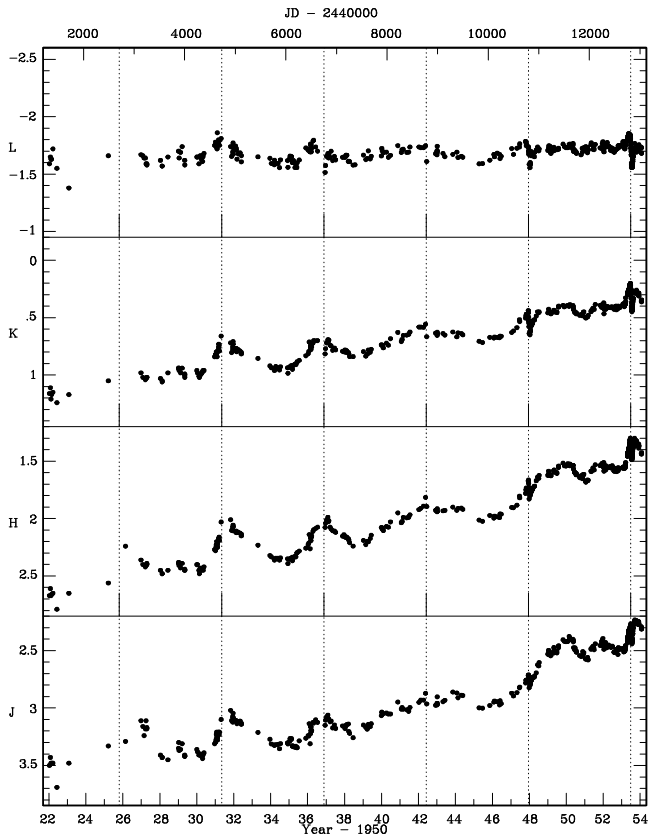}
\caption{\label{jdplot} The $JHKL$ magnitudes for $\eta$ Car over the last
32 years. The dotted lines mark the positions of phase zero calculated
according to the X-ray ephemeris (see section 1).}
\end{figure}

\section{origin of near-infrared flux} 
 Most of the flux from $\eta$ Car, famously, emerges at mid- to far-infrared
wavelengths and comes from dust in the Homunculus emitting over a range of
temperatures (e.g. Smith et al. 2003 and references therein). This thermal
emission is presumed to be reprocessed short-wavelength radiation from the
central star(s). At radio (cm) wavelengths we see optically thick free-free
emission from the stellar wind and from gas ionized by the central source(s)
(e.g. White et al. 1994), while in the visual we see mostly scattered light
from the Homunculus except at very high spatial resolution near the centre
of the nebula, where the reddened stellar wind can be resolved (Morse et al.
1998). The emission line contribution to visual magnitudes is important, but
it seems to originate at some remove from the main continuum source, in the
outer parts of the stellar wind.

The near-infrared continuum has been attributed variously to scattered
light, free-free emission, hot dust and/or the optically thick extended
stellar atmosphere of the central source. The difficulty in settling this
lies, at least partially, with the patchy, variable and uncertain level of
the circumstellar extinction (see section 4.1).
\begin{figure}
\includegraphics[width=8.5cm]{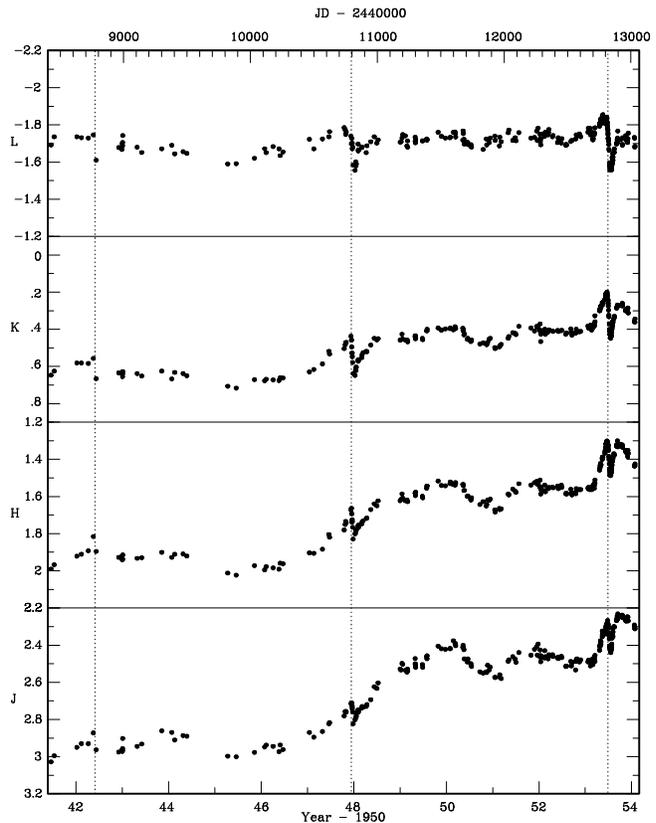}
\caption{As for Fig.~1, but showing only the last two cycles.}
\end{figure}

While images of $\eta$~Car are extended at all wavelengths they reach a
minimum size, or a maximum central concentration, in the near-infrared at
around 2\,$\mu$m (Allen 1989). This is an optical depth effect; at
shorter wavelengths our line of sight does not penetrate far into the dust
and at longer wavelengths emission from that dust dominates what we observe.
Smith \& Gehrz (2000) have conjectured, on the basis of morphological
changes, that most of the near-infrared radiation does not come
directly from the star itself, but from variably illuminated circumstellar
ejecta. It is clear from their images that there is low level continuum
emission scattered from the dust in the Homunculus in the same way as the
optical light is scattered. 

In a detailed analysis of 2 to 12\,$\mu$m images Smith, Gehrz \& Krautter
(1998) estimate the $2\,\mu$m contribution scattered from the bipolar lobes
as 13.4 percent.  Walsh \& Ageorges (2000), on the basis of
$JHK$ polarization measurements and the similarity of the polarization
morphology to that observed in visual bands, also argue that scattering
dominates everywhere except in the central core. However, they find less
polarization at $JHK$ than they expect from an extrapolation of the
polarization at shorter wavelengths and suggest that this might be due to
dilution by non-polarized emission. They consider dilution of the
scattered light by hot dust emission, but reject it because the effect would
be much stronger at $K$ than at $J$, contrary to what is observed. Dilution
by free-free emission would not be strongly wavelength dependent and we
suggest that emission of this kind is a significant contributor, as is
discussed below.

A small component of the flux ($<10$ percent of the total at $K$) must come
from direct line emission (Smith \& Gehrz 2000), including a small
contribution from the narrow emission lines in the Weigelt blobs (these are
bright knots in the ejecta $<0''.3$ from the central star (Smith 2002)).
Lines such as Br$\gamma$ are strongly concentrated towards the core regions
and presumably originate in the extended stellar wind and surrounding HII
region, or just inside the dusty torus (e.g. Smith et al. 1998; Smith \&
Gehrz 2000).

Van Boekel et al. (2003) estimate that in early 2002 about 200 Jy, or almost
half the $2.39\,\mu$m flux, came from a volume with a diameter of 5 mas or
11 AU (at a distance of 2.3 kpc). This would imply a minimum blackbody
temperature of 2300K (assuming no extinction); the temperature would be
higher if allowance were made for plausible extinction. They conclude that
they have spatially resolved the ionized stellar wind. It is clear from
their fig.~1 image (taken on 2 February 2002, van Boekel private
communication) that the rest of the flux in their 1.4 arcsec diameter image,
comes from the immediate circumstellar environment, including the Weigelt
blobs, and from the polar wind extensions to the NW and SE.

It is interesting to compare the image from van Boekel et al. (2003) (which
is actually clearer in the reproduction by Richichi \& Paresce 2003) with
those at similar wavelengths discussed by Smith et al. (1998) and Smith
\& Gehrz (2000), noting that some time-dependent morphological changes must
be expected. In the small area covered by the van Boekel et al. image one
does not see the structure to the NE or SW and therefore has no sense of the
equatorial ``torus" which is prominent at longer wavelengths (Smith et al.
1998). It is not clear if this ``torus" is a low resolution optical illusion
or if it is real and variable (see also Smith et al. 2002). However, the
similarity of the structure seen in fig 4a of Smith et al. (1998) - a
deconvolved $2.16\,\mu$m image to which Br$\gamma$ must be a significant
contributor - and the radio images from Duncan \& White (2003) {\it at
certain phases}, suggests that variable free-free emission is a real feature
of the gas in this region (see below). 

It is not clear whether we need to postulate the existence of any
significant quantities of hot ($T>>400$K) dust in $\eta$~Car. If there is
dust at $T\sim 1000$K then it would have a much greater influence at $L$
than at shorter wavelengths and the fact that the variability at $L$ has
rather different characteristics, both in terms of secular changes and
quasi-periodic variations, may indicate a dominant contribution from dust at
this wavelength. Rigaut \& Gehring (1995) concluded that colours derived
from high spatial resolution images at $JHKL'M$ did indicate the presence of
1000K dust within 100 AU of the central star. Given the results of van
Boekel et al. (2003), the very patchy extinction and the distribution of
line emission, it is not clear that the colours can be interpreted in this
way.  Smith et al. (2003) require a dust component at 550K in the core to
explain a fraction of the 4.8\,$\mu$m flux, and their analysis would not
rule out a small contribution from even hotter dust. The Smith et al. (2003)
dust, at 550K, would contribute about 25 percent of the flux we observe at
$L$.

It seems reasonably certain that free-free emission is a major contributor
to the $JHK$ flux, but it is difficult to establish exactly what proportion
of the emission arises in this way.  There is limited evidence for similar
morphological changes in the 3\,cm (Duncan \& White 2003) and near-infrared
images (Smith \& Gehrz 2000), in the sense that the radio images are clearly
most point-like around phase zero, and near-infrared images may show a
similar tendency (see section 4.3). The fact that there is rather little
correlation between the near-infrared quasi-periodic luminosity variations
and the 3\,cm emission (Duncan \& White 2003; White 2004) ($K$ and 3\,cm
emission seem to have been anti-correlated between 1992 and 1998) must be
attributed to optical depth effects, with the emitting regions being largely
optically thick at 3\,cm and partly optically thin at $2\mu$m.

We can estimate from the 3\,cm flux that there could plausibly be a very
significant contribution at $JHK$, and possibly $L$, from free-free emission
as follows: assuming the emission is partly optically thick, with a power
law spectrum $S_\nu \propto \nu^\alpha$ and a spectral index, $\alpha=+0.6$
(for an unrealistic steady state, isothermal, radially symmetric wind
(Wright \& Barlow 1975)); the free-free flux at $K$ and $L$ would vary
between 230 and 1060\,Jy, when the 3\,cm flux went from 0.7 to 3.2 Jy (1992
to 1998 Duncan \& White 2003). The observed $K$ flux ranges from 320 to
430\,Jy and experiences an uncertain reddening, while that at $L$ is about
1300 Jy. The predicted values will be greater if the power law is steeper,
as it typically is for WR winds, or smaller if the emission has become
optically thin at wavelengths longer than $K$. This serves only to
illustrate that significant contributions from free-free are likely. 

Using the information from van Boekel et al. (2003), Smith \& Gehrz (2000)
and Smith et al. (1998) we estimate that the flux at $K$ measured
through a large aperture (in late 1998) is made up very roughly as: 15
percent light scattered by cool dust, mostly in the Homunculus nebula, 5
percent from emission lines mostly close to the core, 30 percent in the
equatorial ``torus" (free-free), 30 percent in the unresolved core and 20
percent from the region within about 0.7 arcsec of the central core - some
of this will be free-free emission from the polar outflow, but if there is
hot dust present it will be within 0.7 arcsec of the central core and part
of this component.

\section{Infrared Variability}
 The infrared light curves (Fig.~1) show variability on a variety of
timescales that were discussed in Paper III. There is a secular brightening
through all filters which is wavelength dependent, being most pronounced at
$J$ and barely discernible at $L$. There is the, now well established,
``eclipse-like event'', with a period of 2023 days which is most evident in the $L$ data -
where it is deepest and not masked by other variations. Finally there are
quasi-periodic variations which occur on roughly the same time scale as the
events, but which do not seem to be strictly periodic. Despite the wealth
of data none of these variations can be unambiguously explained. They are
discussed below in more detail.

\begin{figure*}
\includegraphics[width=17.6cm]{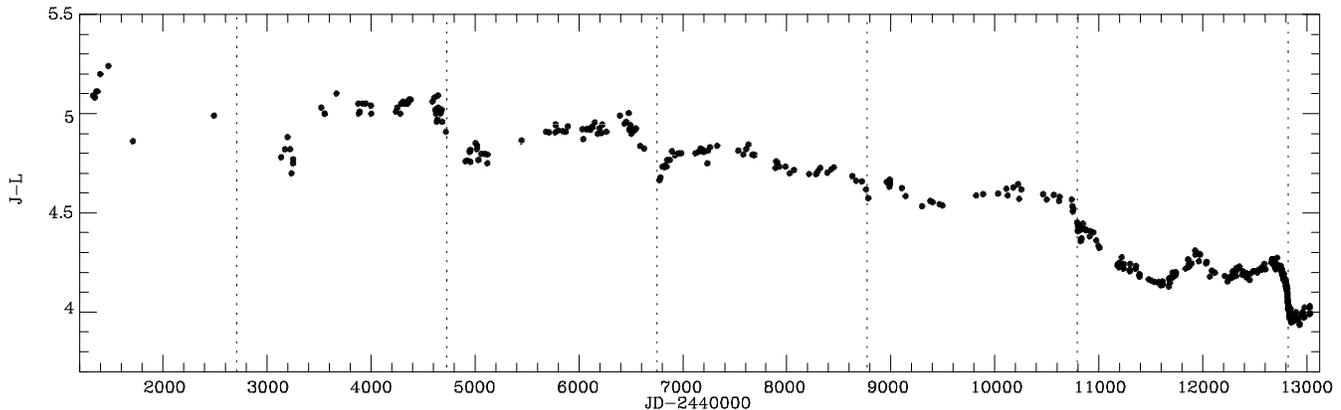}
\caption{\label{jl} The $J-L$ colour as a function of Julian date. The
dotted lines mark the epoch of the X-ray event, as in Figs.~1 and 2.
Note the way the colour changes rapidly during the events and rarely returns
afterwards.}
\end{figure*}

\subsection{Secular Brightening} 
 There is a very clear secular brightening of $\eta$~Car in the $JHKL$
wavebands over the monitoring period of 32 years. Unfortunately our inability
to properly characterize and model the periodic and quasi-periodic
variations (discussed below), which are evident in all wavebands, also 
makes it impossible to describe these secular changes in detail.

\begin{figure}
\includegraphics[width=8.5cm]{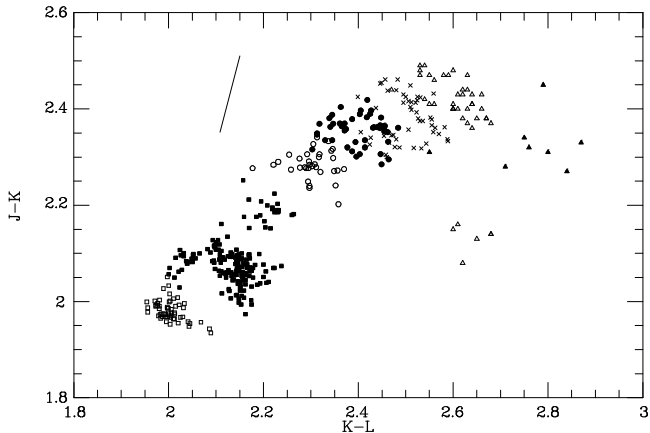}
\caption{\label{jkkl} Two-colour diagram for $\eta$ Car; different symbols
show the various cycles as marked in Figs.~1 and \ref{jl}, going from
earliest to latest as follows: closed triangles, open triangles, crosses,
closed circles, open circles, closed squares, open squares. The line shows
the change in colours that would occur if the interstellar extinction were
to alter by $A_V=1$\,mag (this assumes a normal reddening law which is
probably not applicable to $\eta$ Car; it is shown for illustrative purposes
only).}
\end{figure}

While the changes at $K$ could be consistent with a systematic linear
brightening, it seems likely that at $J$ $\eta$~Car has been brightening
faster over the last 15 than over the previous 15 years, and there can be no
doubt that the overall change in colour has been greater over the last two
cycles than previously. However, for the purpose of the discussion we treat
all trends as linear and Table~\ref{brt} lists the parameters derived from a
simple least squares fit to the magnitudes as a function of Julian Date. The
last column lists the magnitude change over 30 years. A comparison with the
numbers in table~2 of Paper~II suggests that the rate of brightening has
increased significantly, particularly at $J$ and
$L$, but it is important to remember that the exact positions of the end
points affect this simple line fitting. Removing the early points,
which fall off the general trend of colour seen in Fig.~\ref{jkkl}, changes
the values quoted in Table~\ref{brt} by only a small amount.

%
% TABLE 2
%
\begin{table}
\centering
\caption[]{\label{brt} Linear rates of brightening.}
\begin{tabular}{crc}
\hline
band &
\multicolumn{1}{c}{rate}& $\Delta$mag\\
&\multicolumn{1}{c}{mag.day$^{-1}$} & 30\,yrs\\
\hline
$J$ & $-11.2\times 10^{-5}$ & 1.22\\
$H$ & $-10.8\times 10^{-5}$ & 1.18\\
$K$ & $-7.0 \times 10^{-5}$ & 0.77\\
$L$ & $-8.6\times 10^{-6}$ & 0.09 \\
\hline
\end{tabular}
%\end{center}
\end{table}

Davidson et al. (1999) discuss an apparently increased rate of brightening at
near-infrared and visual wavelengths during 1998. This was seen most
dramatically in the HST/STIS observation of the central star (area $0''.1
\times 0''.15$) where the flux at 4000\AA\ increased by 0.83 mag in just 
over one year (see also section 4.3). Examining the infrared changes, in the
context of Figs.~1 and \ref{jl}, and the variations that have taken place in
recent years, it is clear that there have been phases of rapid change; 1998
was one and so was 1981. Of course we have no idea what HST/STIS
observations might have shown in 1981, but the changes of 1998 may be
a common occurrence.

Various possible causes of the secular brightening were discussed in Paper
II. It remains possible, with many caveats, that a reduction of the
extinction due to dust or an increase in flux from the central source or a
combination of the two may be affecting the changes. The secular brightening
is accompanied by a change in the colours, as is illustrated in
Figs.~\ref{jl} and \ref{jkkl}. It is evident from these two diagrams that
the colour changes are more episodic than gradual and that the colours tend
to change discontinuously at the epochs of the event. This suggests that
whatever happens at around these epochs may also be driving the secular
brightening and it would not be consistent with a gradual thinning of the
dust as the major cause of brightening at $JHK$. 

It is difficult to prove that the secular brightening is directly related to
the periodic events, but the magnitude changes illustrated in Figs.~1 and 2
also support that interpretation. At $L$ there is very little secular change
and the quasi-periodic variations always peak just before the eclipse-like
event. At $J$, where the secular changes are at their largest the
quasi-periodic variations very clearly peaked after the eclipse-like event
in both 1998 and 2003/4. This difference is more marked at $J$ than at $H$
where the quasi-periodic variations had their largest amplitude at earlier
epochs. Until we can disentangle the different sources of variability it
will not possible to prove or disprove the link between the secular changes
and the eclipse-like events.

It is clear, from the high spatial resolution HST observations that the
extinction towards the central regions is very patchy (Davidson et al.
1995).  Furthermore, different components of the emission originate in
spatially separate locations and may experience different extinctions. This
applies to different spectral components at the same wavelength; the
continuum, the broad lines and the narrow lines are produced in 
different places. Furthermore, maps at $J$, $K$ and $L$ show slightly
different morphology, suggesting that we are seeing to different depths in
the continuum at these different wavelengths (e.g. Rigaut \& Gehring 1995).
The reddening law for the circumstellar extinction in $\eta$~Car is thought
to be peculiar (Paper II; Davidson et al. 1995 and references therein; Walsh
\& Ageorges 2000), possibly because of an unusual size distribution of the
dust grains. However, the patchy nature of the extinction combined with the
spatial separation of the various emission components, actually make it very
difficult to deduce the reddening law (Hillier \& Allen 1992). Thus, while
the magnitude changes listed in Table~2 are not what we would expect from
the reduction of normal reddening we cannot rule out that explanation
entirely.

Groh \& Damineli (2004) noted secular changes in the emission lines over the
last 11 years, in addition to the variations that take place over the 2023
day period. They record a general weakening of the high- and
intermediate-excitation lines. The equivalent width of the HeI $\lambda$6678
emission has been gradually decreasing, while the absorption component of
this P~Cygni line has been getting deeper. These changes suggest that the
optical depth of the emitting region(s) is gradually increasing and cannot
be attributed to an expansion or thinning of the obscuring dust.
 
Although it is possible to envisage the colour changes, illustrated in
Figs.~\ref{jl} and \ref{jkkl}, as the consequence of grain destruction by
increased ultraviolet flux at around phase zero, the grains would have to be
close to the central star. They would therefore be radiating at a high
temperature and their destruction should have noticeably reduced the $L$
flux. In fact, the secular change in $L$ has been very small, it has been
brightening by less than 0.1 mag in 30 year - so this explanation must be
rejected.

%Said by Davidson et al. 1999 that J produced primarily by free-free and L
%from hot dust located a few hundred au (0.3 arcsec) from star.

%Polomski et al. (1999 AJ 118,2369) may have been a slight brightening at N 
%at the same time as V brightened, not obvious if secular or what.

\subsection{The Eclipse-Like ``Event"}
 First we estimate the interval between events. The 2003 minimum at $K$ was
on JD 2452840.3. In the previous cycle the minimum was between 2450809.6 and
2450824.6, but nearer the latter so we take it to have been on 2450817. The
event in the cycle before was not usefully observed, although minimum must
have been a few days after 2448783 (contrast $J$ and $L$ in Fig.~2 given
that the decline starts sooner at $L$) and for the cycle before that it was
at 2446775.6 or a few days before. These timings are consistent with the
event being a precisely cyclical event and our best estimate for the period
is $2023\pm3$ days. This is almost identical to the period determined from
the X-ray events (Corcoran 2004) and not significantly different from the
2025 days determined from the spectroscopic changes (Groh \& Damineli 2004),
or the $2022.1\pm0.4$ days that Damineli et al. (in preparation) derive from
a more detailed analysis.

\begin{figure*}
\includegraphics[height=20.5cm]{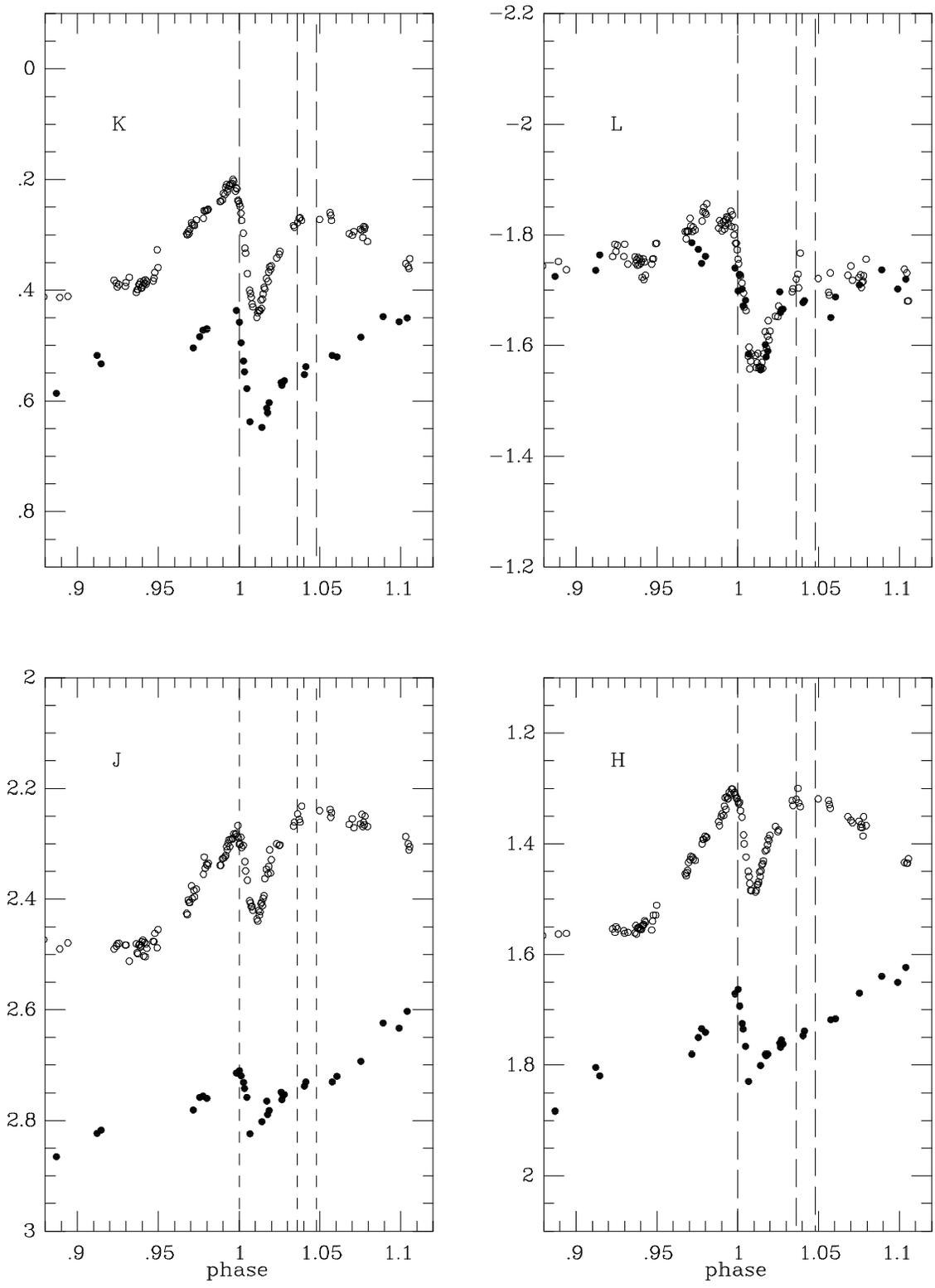}
\caption{\label{4comp} A comparison of the 1998 (solid symbols) and the 2003
(open symbols) events at the four wavelengths. The dashed lines are taken
from the X-ray event (Corcoran 2004) and mark the phases of the start and
end of X-ray minimum and fourth contact.}
\end{figure*}

Fig.~\ref{4comp} allows us to compare the 1998 and 2003 events and to make a
comparison with the X-ray light curve (Corcoran 2004). The two events
naturally separate in magnitude at $JHK$ because of the long-term trend. At
$L$, where the long-term trend is negligible over 5 years, the differences
in shape and amplitude are minor. It is of course very difficult to be
certain whether changes at any wavelength are linked to the event or to some
other, possibly unrelated, variations.
During the event the hard X-ray emission drops to one percent of the flux
before minimum (Hamaguchi et al. 2003; Corcoran 2004) and stays close to
that level for about 70 days. The infrared event is preceded by a
brightening at all wavelengths (but see section 4.3), as is the X-ray event. 
In marked contrast to the X-ray light curve, the infrared event involves
flux decreases of only 10 to 24 percent and is flat bottomed only at $L$ and
only for about 20 days.  While the oscillations in the X-ray flux make it
difficult to pinpoint the time of ingress, it is clearly earlier than it is
at infrared wavelengths. Within the near-infrared light curves ingress starts
earliest at $L$ and latest at $J$  where it more or less coincides with the
well-defined start of X-ray minimum; the delay between $L$ and $J$ is about
7 days. The $K$ minimum occurs at X-ray phase 0.011.

%
% TABLE 3
%
\begin{table}
\centering
\caption[]{\label{eclipse} Comparison of eclipse-like events.}
\begin{tabular}{ccccc}
\hline
band & \multicolumn{2}{c}{depth} & \multicolumn{2}{c}{2003/1997}\\
& 1997 & 2003 & max & min \\
\hline
$J$ & 10\% & 14\% & 1.49 & 1.43 \\
$H$ & 15\% & 15\% & 1.39 & 1.38 \\
$K$ & 18\% & 20\% & 1.24 & 1.20 \\
$L$ & 19\% & 24\% & 1.06 & 1.00 \\
\hline
\end{tabular}
%\end{center}
\end{table}
According to the data of Fern\'andez Laj\'us et al. (2003) the fading at
visual wavelengths, $BVRI$, started later still (JD\,2452826 at $V$), about
10 days after $J$. It should be borne in mind that the $BVRI$ measurements
are much more strongly influenced by scattered light from the Homunculus
than are $JHKL$. The optical high-excitation emission lines disappear around
phase zero (Groh \& Damineli 2004). Note that these originate in an extended
region around the central star, not in the stellar wind itself (Smith et al.
2000).  

The data in Table~\ref{eclipse} allows us to compare the depths of the last
two events. The second and third columns give the depth at minimum,
expressed as a percentage of the flux just before the event. The next two
columns are the flux ratio, 2003 divided by 1997/8, for the pre-event
maximum (column 5) and the event minimum (column 6). The errors are about
two percent on all the numbers. Taking these at face value we see that the
event stayed the same or deepened slightly between 1997/98 and 2003. This
means that whatever has brightened during that interval decreases in
approximately the same proportion as or slightly more than before. Although
it is clear that the event is about 10 percent deeper at $L$ than at $J$,
and of intermediate depth in the intervening wavebands, this may be a little
deceptive. The time of peak brightness is wavelength dependent (possibly
because of the secular trend) and $J$ does not reach its peak until after
the event, as in earlier cycles. Obviously the eclipse-like event would be
deeper at $J$ if we measured it with respect to this peak rather than to the
maximum before the event.

The 1998 event showed a two step egress (Paper III, figs.~4 and 7), the
shallow part of which extended in phase from about 0.01 to 0.09, ending at
about the same time as the low excitation phase of the emission lines. The
2003 event also shows a two phase egress, but the shallow part is brief,
extending only from phase 0.01 to 0.036. That is, it ends about the same time
as the X-ray event comes out of its minimum.  It was clear from the
historical analysis of Paper III that the low excitation event does not
cover the identical phase every time, but we have almost no information
about infrared flux levels for earlier events. It will be interesting to see
if the 2003 return of the high excitation spectrum correlates with the
infrared light curve in the same way as it did in the previous event.

Gull et al. (2003) describe changes in the ultraviolet as observed at high
spatial and spectral resolution with HST/STIS during the critical phases.
Visibility of the central source dropped, but not uniformly throughout the
ultraviolet, so they suggest that the fading was caused by multiple line
absorptions both in the source and in the intervening ejecta. High
excitation lines diminished or disappeared while lines originating from
lower levels strengthened. Observations with FUSE also show striking
wavelength-dependent absorptions during the event (Iping et al. 2003).
Longward of 1100\AA\ the overall flux dropped by 10 to 30 percent, but
shortward of this wavelength there were intervals with no decrease in flux
at all. Thus dust absorption must play a negligible role in producing the 
events.

Martin et al. (2004) describe HST/STIS data relevant to the event which is
particularly interesting as it isolates the behaviour of the central star.
They find that H$\alpha$ fades at phase 0.95, much earlier than even the
X-rays and at about the same time as the pre-event brightening starts at
$JHKL$, but brightens again at phase 1.04. The continuum at 6770~\AA\
brightens as H$\alpha$ fades. The $V$ continuum does not show any
event-related change at all, in obvious contrast with ground-based $V$
observations which encompass the whole Homunculus.

\subsection{Quasi-periodic Variations}
 In Paper II we noted quasi-periodic variations in the $JHKL$ data for 
$\eta$ Car covering 20 years up to 1994; their characteristic time scale was
about 5 years. Marking the 2023 day cycles in Fig.~1 guides the
eye to see the changes at this period. It is particularly clear in the $H$
light curve that peaks occur at around the time of events at five of the
six marked intervals. A Fourier analysis of these data, after removing the
long-term trend (assumed to be linear), indicates primary periods of 1961
days at $L$, 2091 days at $K$, 2074 days at $H$ and 2044 days at $J$. Given
that the variations are not sinusoidal, and that there are other changes on
different time scales, these results are consistent with an underlying
period of 2023 days. However, we would probably not deduce strict
periodicity for these particular variations if it were not for the 2023 day
periodicity of the eclipse-like events.

The peak-to-peak amplitude in magnitudes of the quasi-periodic variations is
0.2 to 0.3 at $J$, 0.35 to 0.45 at $H$, 0.2 to 0.3 at $K$ and 0.1 to 0.2 at
$L$; the maximum amplitudes are good to $\pm 0.03$ mag. It is at $L$ that
these variations are at their most repeatable (see Paper III), with the
maximum always occurring just before the eclipse-like minimum. We might
estimate that roughly one-third of the $L$ flux originates from warm dust
emission (see section 3). Subtracting this dust flux leaves a component with
an amplitude of about 0.3 mag - comparable to values measured at $J$ and $K$.
Unfortunately a meaningful comparison of the spectral energy distribution of
these quasi-periodic variations cannot be made without removing the
contributions from other sources of emission - which cannot be quantified as
yet.

At $J$, $H$ and $K$ the peak that occurs just before the event minimum seems
to form part of these quasi-periodic variations. Sometimes this peak is very
close to the broad maximum, as in 1981 and 1986/7, at other times it is very
early on, as in 1997/8. The most recent variations may have increased in
amplitude at $J$ relative to $H$, but it is difficult to be certain because
of the poorly characterized secular changes.

Smith \& Gehrz (2000) compare two high-spatial-resolution $K$ measurements
of $\eta$~Car made in May 1995 (JD\,$2449852\pm15$, phase 0.53) and on 6
September 1998 (JD\,2451063, phase 0.13). They find that the flux within a
central $1''$ aperture increases by a factor of 1.93 between 1995 and 1998
and its distribution becomes more point-like in 1998 when it is brightest.
The data in our Table 1 suggests that the integrated $K$ flux goes up by a 
factor of 1.27 over the same period (note that Smith \& Gehrz find that the
integrated flux changes by only 13 percent from their images; the difference
between their measurement and ours is only in the value from the 1998 NICMOS
measurement). Davidson et al. (1999) discuss the brightening at other
wavelengths (see also section 4.1) and note that the central star brightened
in the optical and near-ultraviolet. The brightening seems to have been
nearly wavelength independent.

van Genderen, de Groot \& Sterken (2001) find luminosity peaks in the visual
data following the same kind of pattern that we see in Fig.~1, although with
amplitudes of only one or two tenths of a magnitude. They suggest that these
may be a consequence of a binary companion triggering ``S Doradus events",
i.e. mass loss, from the primary and the mass moving into a disk or torus.
They also discuss problematic aspects of this interpretation.

The radio flux also shows variability, by more than a factor of three, on
this same time scale (Duncan \& White 2003), although there does not seem to
be much correlation between the intensities of the radio and near-infrared
variations as was discussed in section 3. Duncan \& White also interpret the
changes as a consequence of the tidal transfer of material from the primary
into a disk. The radio image is much more compact when the radio emission is
faint - possibly it is only the extended component, with a diameter of 4 to
5 arcsec at maximum, that varies significantly and that it does so because
of changing levels of ionizing radiation, as suggested by Duncan et al.
(1995).

Damineli's (1996) discovery of the inverse correlation of the equivalent
width of the HeI 1.083\,$\mu$m line with the $H$-band flux led to a dramatic
change in our thinking on $\eta$~Car.  Like many other aspects of the
variability it is most easily understood as the consequence of changes in
the optical depth of the emitting region(s).

It is vital that $\eta$~Car be monitored at high spatial resolution in the
near-infrared if further progress is to be made in interpreting the various
changes and their relationship to the central star(s).

\section{Discussion}
 In view of the emphasis that Smith et al. (2000) put on the non-periodicity
of the infrared flux variations (last paragraph of their section 4.2) it is
important that we clarify this point. In section 4.2 we derived a period for
the ``eclipse-like events" of $2023\pm3$ days; this is, as far as current
information allows us to determine it, a ``precise period". In contrast, the
broad peaks seen in the $JHKL$ light curves over the last 30 years (see
section 4.3) are quasi-periodic.

The similarity of the timescale of the quasi-periodic variations and the
period between events suggests that the underlying cause of these phenomena
is the same and that they are both orbitally modulated variations.
Furthermore, the secular variations are characterized by discontinuous colour
changes which occur around phase zero in the 2023 day cycle, which suggests
that they too may be driven by something associated with the orbital period.

The morphology of the X-ray light curve during the 1997/8 event was
explained by Corcoran et al.  (2001) in terms of a binary model with
periastron passage at phase zero. The X-rays are produced by shock-waves in
colliding stellar-winds and the specifics of the light curve can be
reproduced if the mass-loss from $\eta$ Car increases from $\rm 3\times
10^{-4}M_{\odot}yr^{-1}$ by a factor of about 20 for a period of about 80
days following periastron. During the event the X-rays are absorbed in the
dense wind resulting from the enhanced mass loss. Zanella et al. (1984)
first suggested that the low-excitation spectroscopic phase and enhanced
near-infrared emission might be caused by a shell ejection; although we note
that they attributed the increased near-infrared flux to dust formation
which is contrary to our results. When the 5.5\,yr binary period started to
look plausible various people conjectured that the close passage of the
secondary in an elliptical orbit might trigger increased mass-loss at
critical phases (e.g. Davidson 1999; Ishibashi et al. 1999).

Other evidence, cited in section 4.2, points to enhanced ultraviolet
line opacity during the event, resulting in decreased excitation of emission
lines at many wavelengths. Surprisingly there is no evidence for changes in
the temperature or luminosity of the visual continuum from the central
source and it is possible that the optical depth is sufficiently high that
what is observed at high spectral resolution is well away from where the
action is taking place. We have insufficient information to be certain what
is causing the event observed at $JHKL$, but we might conjecture that
free-free emission, coming from a rather broad volume around the central
source, will greatly decrease if its source of ultraviolet excitation is
quenched when the secondary is enveloped in the material pulled from the
primary or from the passage of the secondary through dense pre-existing
circumstellar material. 
 
In this picture the quasi-periodic changes would be explained in a similar
way to the radio variations (Duncan \& White 2003) if the secondary star is
largely responsible for ionizing the HII region surrounding $\eta$~Car.
Orbital modulation of the source of ionizing radiation as it moves in and
out of the denser parts of the circumstellar disk will drive the
quasi-periodic variations.  Much of the nebula must be optically thick at
radio wavelengths but not in the near-infrared. 
 
Thus the secular variations may also be the consequence of enhanced
free-free emission as the newly ejected mass joins the material around
$\eta$ Car and emits at $JHK$ wavelengths. It may well be possible to use
the information from the multi-wavelength campaign to establish where the
material is building-up. We might speculate that if conditions become
suitable for dust formation in these ejecta then radiation pressure could
cause the dust and gas to be expelled with great efficiency. 

\section*{Acknowledgments} We are very grateful to Lisa Crause and Dave
Laney for making some of the $JHKL$ measurements and to John Menzies for a
critical reading of the manuscript. We also thank the following for their
comments and suggestions on the preprint: Mike Corcoran, Augusto Damineli,
Kris Davidson, Roberta Humphreys, John Martin and an anonymous referee.

\end{document}